\documentclass[aps,prb,twocolumn,superscriptaddress]{revtex4}
\usepackage{graphicx} 
\usepackage{dcolumn}
\usepackage{bm}

\begin{document}

\title{Femtosecond Demagnetization and Hot Hole Relaxation in Ferromagnetic \nobreak{GaMnAs}}

\author{J.~Wang}
\altaffiliation{Present address: Lawrence Berkeley National Laboratory, Berkeley, CA}
\email[]{JWang5@lbl.gov}
\affiliation{Electrical and Computer Engineering Department, Rice University, Houston, Texas 77005, USA}

\author{{\L}.~Cywi{\'n}ski}
\altaffiliation{Present address: Condensed Matter Theory Center, Department of Physics, University of Maryland, College Park, MD}
\email[]{lcyw@umd.edu}
\affiliation{Department of Physics, University of California, San Diego, La Jolla, California 92093, U.S.A.}

\author{C.~Sun}
\affiliation{Electrical and Computer Engineering Department, Rice University, Houston, Texas 77005, USA}

\author{J.~Kono}
\thanks{Corresponding author}
\email[]{kono@rice.edu}
\affiliation{Electrical and Computer Engineering Department, Rice University, Houston, Texas 77005, USA}

\author{H.~Munekata}
\affiliation{Imaging Science and Engineering Laboratory, Tokyo Institute of Technology, Yokohama, Kanagawa 226-8503, Japan}

\author{L.~J.~Sham}
\affiliation{Department of Physics, University of California, San Diego, La Jolla, California 92093, U.S.A.}

\date{\today}

\begin{abstract}
We have studied ultrafast photoinduced demagnetization in \nobreak{GaMnAs} via two-color time-resolved magneto-optical Kerr spectroscopy.  Below-bandgap midinfrared pump pulses strongly excite the valence band, while near-infrared probe pulses reveal sub-picosecond demagnetization that is followed by an ultrafast ($\sim$1 ps) partial recovery of the Kerr signal.  Through comparison with \nobreak{InMnAs}, we attribute the signal recovery to an ultrafast energy relaxation of holes.  We propose that the dynamical polarization of holes through $p$-$d$ scattering is the source of the observed probe signal.  These results support the physical picture of femtosecond demagnetization proposed earlier for \nobreak{InMnAs}, identifying the critical roles of both energy and spin relaxation of hot holes.
\end{abstract}


\maketitle

\section{Introduction}
Ultrafast studies of collective magnetic phenomena can provide fundamental insights into
non-equilibrium processes of correlated spins as well as extend the speed limits of information recording technology.\cite{Stanciu_PRL07}
The discovery of photoinduced sub-picosecond demagnetization in Ni\cite{Beaurepaire_PRL96} has triggered intense interest in spin dynamics
in metallic and insulating magnets.\cite{Zhang_TAP02,Koopmans_TAP03,Kimel_PRL02,Wang_PRL05,Kimel_Nature04, Kimel_Nature05}  Recently, femtosecond demagnetization,
including a total quenching of ferromagnetic order in less than a picosecond,  has been observed in \nobreak{InMnAs},\cite{Wang_PRL05} which
belongs to the family of (III,Mn)V ferromagnetic semiconductors.
We have previously proposed\cite{Wang_PRL05,Wang_JPC06,Cywinski_PRB07} that demagnetization in (III,Mn)V materials occurs
through $sp$-$d$ spin scattering between hot holes and localized Mn spins.  Here we present a further experimental
study of \nobreak{GaMnAs} which supports this physical picture.

\begin{figure}
\includegraphics [scale=0.45] {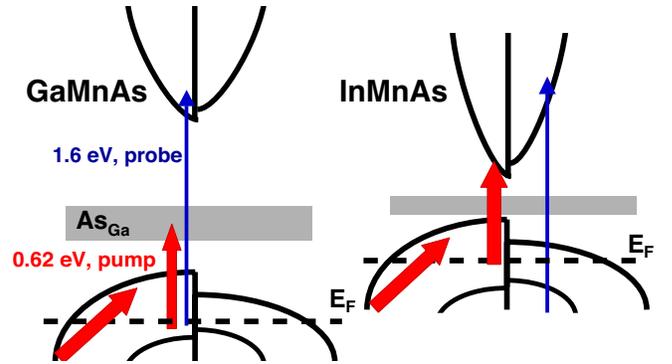}
\caption{(color online) Optical transitions involved in the present experiments on \nobreak{GaMnAs} (left) and \nobreak{InMnAs} (right) in the ferromagnetic state where the conduction band (CB) and valence band (VB) are spin-split.  In \nobreak{GaMnAs}, below-bandgap pump pulses at 0.62~eV (thick solid arrows) create holes in the VB via transitions from below the Fermi level to the midgap states (e.g., As$_{\text{Ga}}$ antisite defect band centered $\sim$0.5~eV above the top of the VB), and excite the
pre-existing hole population by intervalence band transitions allowed through the relaxation of the $k$-selection rule.
In \nobreak{InMnAs}, the photoelectrons in the CB are also created through direct interband transitions.  The magnetization changes are probed via Kerr rotation by reflected, time-delayed pulses at 1.6~eV (thin blue arrows). }
\label{typical}\end{figure}

Carrier-mediated ferromagnetism in (III,Mn)V semiconductors originates\cite{Dietl_Science00} from the mutual interaction of delocalized hole spins ($\textbf{\textit{s}}$) and localized Mn spins ($\textbf{\textit{S}}$).  The natural division into carrier (hole) and spin (Mn ions) subsystems, as well as the simple form of the effective magnetic interaction Hamiltonian between them ($\sim\! \beta \textbf{\textit{S}} \cdot \textbf{\textit{s}}$, where $\beta$ is the $p\!-\!d$ exchange constant), allow for a more transparent treatment of demagnetization than in transition metals.\cite{Koopmans_PRL05}  Photoinduced femtosecond demagnetization in \nobreak{InMnAs}\cite{Wang_PRL05} has been well described by the ``inverse Overhauser effect,'' i.e., the spin-flip scattering between holes and Mn spins enhanced by a high effective temperature of photoexcited carriers.\cite{Cywinski_PRB07}  For this model to explain the magnitude of observed demagnetization, the spin relaxation time of hot holes has to be very short
($\sim$10~fs) so that each hole can participate in multiple spin-flip scatterings\cite{Wang_PRL05,Akimov_PRB06} within a relevant time-scale (a picosecond).
Furthermore, the fact that the fast demagnetization process terminates in less than a picosecond was assumed to be connected with the energy relaxation of excited carriers.  However, the ultrafast energy and spin relaxation of strongly excited holes, and its correlation with femtosecond demagnetization, have not been explicitly addressed experimentally until now.

Here we report ultrafast demagnetization and hot hole relaxation in \nobreak{GaMnAs} via two-color time-resolved magneto-optical Kerr effect (MOKE) spectroscopy. The midinfrared (MIR) pump pulses strongly excite the pre-existing hole population, without creating any photoelectrons in the conduction band.  Near-infrared (NIR) probe pulses reveal ultrafast demagnetization followed by a partial recovery of the MOKE signal with a time constant $\sim$1~ps.  This ``overshoot'' behavior, absent in \nobreak{InMnAs}, indicates the presence of dichroic bleaching\cite{Koopmans_PRL00}
(Pauli blocking of optical transitions) and provides a direct measurement of the energy relaxation of hot holes. The initially highly excited hole population loses its excess energy in less than a picosecond, leading to the rapid termination of the demagnetization process.
We argue that the presence of the dichroic bleaching effects in \nobreak{GaMnAs}, as opposed to \nobreak{InMnAs}, comes from the fact that our probe pulse ($\hbar \omega \! = \! 1.6$~eV) is tuned close to the peak of magneto-optical response of \nobreak{GaMnAs},\cite{Ando_PRL08} and the relevant transitions involve the states close to the Fermi energy. The population of these states is strongly perturbed by the pump.
We discuss in detail the origin of the probe response overshoot and propose that it originates from a transient dynamical polarization of holes, which is a direct consequence of the ``inverse Overhauser'' mechanism of demagnetization.  Finally, the fact that we have not observed any optical orientation effects of holes in \nobreak{GaMnAs} excited with a circularly polarized MIR pump, shows that photoexcited holes relax their spins on a time scale much shorter than the pulse duration ($\sim \! 100$~fs), as it is required for the theoretical model to work.

\begin{figure}
\includegraphics [scale=0.6] {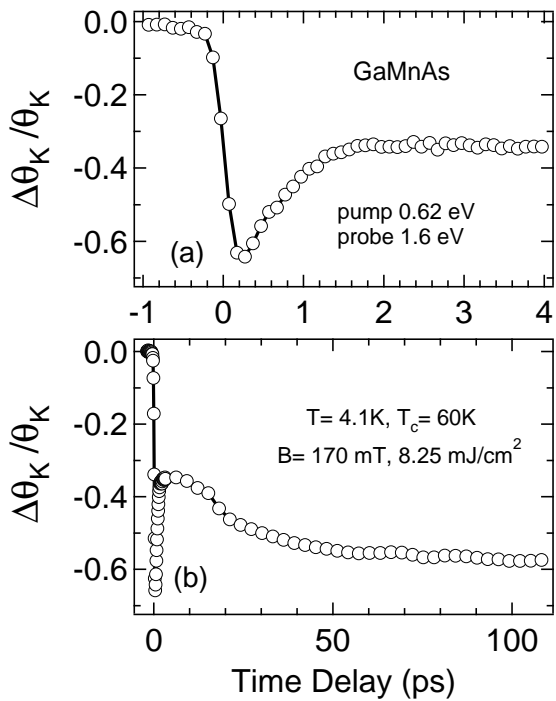}
\caption{(a) First 4~ps of $\Delta \theta_{\rm K}(t)$ for \nobreak{GaMnAs}. (b) $\Delta \theta_{\rm K}(t)$ up to $t=$~120~ps.  The initial signal is negative (i.e., demagnetization) and very fast, followed by a partial recovery of the MOKE angle with a time constant of $\sim$1~ps, and a slow, continuous decrease up to $\sim$100~ps.}
\label{mag-dep}\end{figure}

\section{Experimental results}
The samples studied were a 50-nm Ga$_{0.974}$Mn$_{0.026}$As layer on 1-$\mu$m InGaAs and 300-nm GaAs layers, and a 25-nm In$_{0.87}$Mn$_{0.13}$As layer on a 820-nm GaSb layer. Both were grown  by low-temperature molecular beam epitaxy (LT-MBE) on a semi-insulating GaAs (100) substrate.  The Curie temperature ($T_{\rm c}$) was $\sim$50~K for the GaMnAs sample and $\sim$60~K for the InMnAs sample, and the magnetic easy axes of both samples were parallel to the growth axis due to tensile strain.  In LT-MBE-grown ferromagnetic \nobreak{GaMnAs} and \nobreak{InMnAs} there is no clear bandgap in optical absorption, which is non-zero for practically all energies.\cite{Singley_PRB03,Hirakawa_PE01}
This strong ``band tailing'' is believed to be due to defect states present in the gap, such as As$_{\text{Ga}}$ antisites and Mn interstitials, and $k$-selection rule breaking by disorder.\cite{Szczytko_PRB01,Dziatkowski_PRB06}
As shown in Fig.~1 (left), the MIR pump with 140~fs pulse duration was tuned to 0.62~eV, exciting the valence band in two distinct ways. The first is the generation of more holes by exciting the valence band electrons from below the Fermi level into the localized midgap states, such as the As$_{\text{Ga}}$ antisite level located $0.5$~eV above the top of the valence band.\cite{Lodha_JAP03} The second is the inter-valence band excitation (promoting electrons from below to above the Fermi level in the valence band), allowed by the breaking of the  $k$-selection rule mentioned before. In this process the number of holes does not change, but their temperature is increased. In neither of these processes any conduction band electrons\cite{Wang_PRB05,Sanders_PRB05} are created (unlike the case in Refs.~\onlinecite{Oiwa_PRL02,Mitsumori_PRB04,Kimel_PRL04}), which allowed us to study ultrafast demagnetization dynamics induced  exclusively  by the hot holes.  The magnetization changes were probed via MOKE rotation\cite{Wang_JPC06} through the reflection of a time-delayed NIR pulse with 140 fs pulse duration.  The probe
was tuned at 1.6~eV, which gives a static MOKE angle $\theta _{\rm K}$ of 4~mrad.  The large pump fluences used, $\simeq 8$~mJ/cm$^{2}$, ensured that we detected pure demagnetization dynamics, and not the other subtle transient magnetic effects manifested at low-pump fluences, e.g., photoinduced enhancement of magnetization.\cite{Wang_PRL07,Wang_SPIE08}

Typical time dependence of photoinduced MOKE-angle change, $\Delta\theta_{\rm K}/\theta_{\rm K}$, is shown in Fig.~2 for (a)~the first 4~ps and for (b)~time delays up to 120~ps.  Distinct temporal regimes can be identified: i)~an initial ($<$200~fs) fast reduction of $\Delta\theta_{\rm K}/\theta_{\rm K}$, ii)~a partial recovery of the initial ``overshoot'' with a time constant of 1~ps, and iii)~a much slower further decrease of magnetization occurring on the time-scale of $\sim$100~ps.  Process iii) corresponds to the standard thermal demagnetization via heat transfer from phonons (spin-lattice scattering).\cite{Kojima_PRB03,Wang_PRL05}  A final recovery towards equilibrium appears much later with a time constant of $\sim$~2.5~ns (data not shown).

The magnetic field ($B$) and temperature ($T$) dependences of $\Delta\theta_{\rm K}(t)$ are shown
in Figs.~3(a) and 3(b), respectively.  In Fig.~3(a) it is seen that the sign of $\Delta\theta_{\rm K}$ changes
when the direction of the field is reversed, and the sign of $\Delta\theta_{\rm K}/\theta_{\rm K}$ is always negative (i.e., {\it demagnetization}).  No $\Delta\theta_{\rm K}$ is observed at zero field, indicating that without an external field the magnetization direction is not defined between pump pulses
(separated by 1~ms) since each pump pulse completely demagnetizes the sample.  From this information, using the method described in Ref.~\onlinecite{Cywinski_PRB07} we can estimate the effective hole temperature right after photoexcitation, $T_{\rm h}$, to be $\geq$~700~K for a hole density ($p$) of $\approx$~10$^{20}$~cm$^{-3}$.  Furthermore, the data in Fig.~3(b) shows that $\Delta\theta_{\rm K}$ decreases drastically as $T$ approaches $T_{\rm c}$ and is indeed absent above $T_{\rm c}$.
The overshoot disappears close to $T_{\rm c}$, e.g., as seen in the 45~K trace, and $\Delta\theta _{\rm K}$ is {\em completely absent} above $T_c$. The fact that the magnitude of the overshoot decays with increasing temperature faster than the ``saturated'' demagnetization at $\sim$1~ps time delay is an important clue to the mechanism behind this feature in the probe response. We discuss it further in Section~\ref{sec:bleaching}.

\begin{figure}
\includegraphics [scale=0.6] {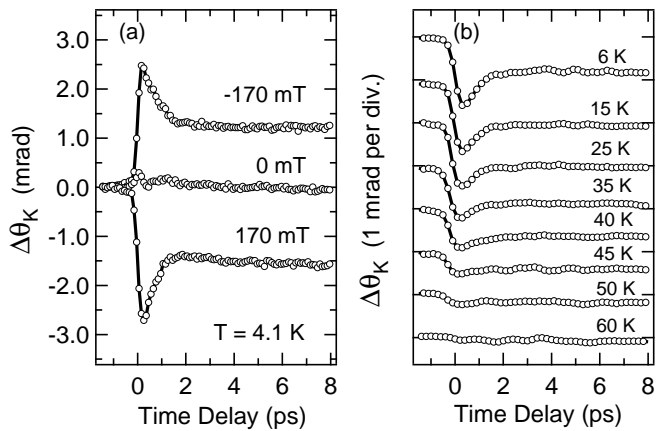}
\caption{(a) Magnetic-field-dependence of $\Delta\theta_{\rm K}(t)$.  (b) Temperature-dependence of $\Delta\theta_{\rm K}(t)$ from 6~K to 60~K.  The signal decreases quickly with increasing temperature and is absent above the Curie temperature ($\sim$50~K).}
\label{power} \end{figure}

\begin{figure}
\includegraphics [scale=0.6] {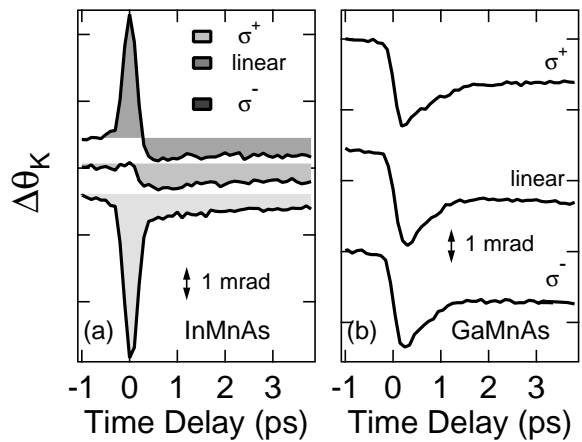}
\caption{Photo-induced MOKE dynamics for (a) \nobreak{InMnAs} and (b) \nobreak{GaMnAs} under pumping with circularly ($\sigma^+$ and $\sigma^-$) and linearly polarized radiation.  The pump and probe photon energies are the same as those in other figures., and the magnetic field is $B \! > \! 0$.}
\label{polarization}\end{figure}

The MOKE responses of \nobreak{InMnAs} and \nobreak{GaMnAs} for circularly-polarized pump excitation are qualitatively different.  Figure~4(a) clearly shows that the initial transients in \nobreak{InMnAs} strongly depend on the pump polarization.  They disappear within 1~ps, similarly to the transient reflectivity results,\cite{Wang_JPC06} which are dominated by free photoelectrons.  From this we infer that these transients are due to the spin-polarized electrons created in the conduction band, and their decay is due to electron trapping by defects, a non-radiative process characteristic for LT-grown materials.\cite{Wang_JPC06}  However, in the case of \nobreak{GaMnAs} pumped with 0.62~eV light, no free photoelectrons are created. The fact that there is no trace of pump-polarization dependence in the probe response [see Fig.~4(b)] strongly suggests that the optical orientation of photoholes is very quickly destroyed by hole spin relaxation, much shorter than the pulse duration used ($\sim$~140~fs).  This is consistent with
the theoretical simulations\cite{Cywinski_PRB07} showing that fast ($\sim$10~fs) spin relaxation of holes is necessary for explaining the observed magnitude of demagnetization.

\section{Interpretation and theoretical modeling}
The magneto-optical response of (III,Mn)V semiconductors below $T_{\rm c}$ is caused by spin splitting of bands due to the ferromagnetic order, as illustrated in Fig.~1.  The optical selection rules arising from spin-orbit coupling in the valence band of III-V semiconductors then lead to different optical responses to the two circular polarizations of light $\sigma_{\pm}$.
Within the $sp$-$d$ model of ferromagnetism,\cite{Dietl_Science00,Dietl_PRB01} the band splitting $\Delta$ is proportional to the average localized spin.
In the static (equilibrium) case, only this band splitting in the ground state is probed, and the MOKE response is proportional to the macroscopic magnetization of the Mn spins.  In time-resolved measurements with {\em strong photoexcitation} discussed here, holes excited by the pump can influence the magneto-optical response in other distinctly different ways.  Besides the splitting of bands, the MOKE signal is also directly influenced by state-filling,\cite{Koopmans_PRL00} provided that the probe light couples to the states whose occupation was strongly modified by the pump.  Therefore, a selective pump-probe scheme can provide an opportunity to study the ultrafast relaxation dynamics of hot carriers.  For example, the previously discussed pump-polarization dependent transients in \nobreak{InMnAs} arise from the creation of spin-polarized photoelectrons in the conduction band, leading to the bleaching of transitions induced by the part of the probe beam co-polarized with the pump.

One of our main observations for \nobreak{GaMnAs} is the initial prominent overshoot behavior, which is not seen in \nobreak{InMnAs}.\cite{Wang_PRL05}  This is clearly visible in the $\Delta\theta_{\rm K}(t)$ traces for a linearly-polarized pump (center) in Figs.~4(a) and 4(b).  In \nobreak{InMnAs} the transitions away from the $\Gamma$ point (originating below the quasi-Fermi level of strongly excited valence band electrons) contribute to the optical response at 1.6~eV.  Due to the very small effective mass in the conduction band of \nobreak{InMnAs} the thermalized photoelectrons can still bleach the probe transitions, but it is plausible that most of the initial states in the valence band contributing to the probe response are far away from those affected by the pump excitation, as illustrated in Fig.~1 (right).  Consequently, apart from the electron-related transient, the MOKE signal in \nobreak{InMnAs} arises from the changes in average Mn polarization.\cite{Wang_PRL05}  On the other hand, in the case of \nobreak{GaMnAs} (see Fig.~1, left), the probe frequency was tuned near the peak of the static MOKE spectra of \nobreak{GaMnAs}.\cite{Beschoten_PRL99,Lang_PRB05,Ando_PRL08}  The transitions responsible for this peak were most commonly assumed to involve final states in the conduction band,\cite{Szczytko_PRB01,Dietl_PRB01} but in recent experiments other possibilities were discussed.  Final states in the dispersionless level close to the bottom of the conduction band were proposed.\cite{Lang_PRB05}  It has also been argued that most of the magnetooptical spectrum of \nobreak{GaMnAs} at $\sim$1.6~eV is dominated by transitions involving the intra-gap states.\cite{Ando_PRL08,Picozzi_PRB06}  Our findings support  these interpretations, as our calculations of magnetooptical spectra of \nobreak{GaMnAs} using the $\mathbf{k}$$\cdot$$\mathbf{p}$ band model\cite{Dietl_PRB01} agree qualitatively with the published measurements\cite{Beschoten_PRL99,Ando_PRL08} only for a very low hole concentration, $p \sim$~10$^{19}$~cm$^{-3}$, which disagrees with the critical temperatures of most of the measured samples.  In any case, the probe couples to electronic transitions for which the initial states are in the valence band, close to the Fermi level. These are exactly the states whose population is strongly disturbed by the pump. Taking into account that the holes relax their spin very quickly, we expect prominent, pump-polarization-independent dichroic bleaching in \nobreak{GaMnAs}.  This expectation is confirmed by results shown in Fig.~4(b).

\subsection{Sources of dichroic bleaching} \label{sec:bleaching}
There are two possible sources for the dichroic bleaching signal observed in \nobreak{GaMnAs}.  The first is a simple blocking of transitions by photoholes added to the valence band, which diminishes the absorption $\alpha_{\pm}$ for both circular polarizations.  Although $\theta_{\rm K}$ is a complicated function of the optical constants of the material (especially in the relevant case of reflection from a layered material), it is generally proportional to $\alpha_{+}-\alpha_{-}$. When the two absorption coefficients are down-scaled by the same factor, the MOKE signal is suppressed.

The second source of dichroic bleaching is the dynamical polarization of holes created in the process of demagnetization. The following scenario  assumes that the Kerr response of the probe corresponds not to the average band splitting (proportional to the average Mn spin), but to the average polarization of hot holes in the valence band.

When the holes are very hot, the enhanced rate of $p$-$d$ scattering with the Mn spins leads to the transfer of angular momentum from the localized spins to the holes. Unless the hole spin relaxation is instantaneous, the nonequilibrium (dynamical) spin polarization of holes is created, i.e., their average spin ceases to be proportional to the average Mn spin. This is described by the equation for the dynamics of the average hole spin $\langle s^{z}(t) \rangle$
\begin{equation}
\frac{d}{dt} \langle s^{z}(t) \rangle = - \frac{n_{\text{i}}}{p} \frac{d}{dt} \langle S^{z}(t) \rangle - \frac{  \langle s^{z}(t) \rangle - s_{0}(t) }{\tau_{\text{sr}} } \,\, , \label{eq:dsdt}
\end{equation}
in which $n_{\text{i}}$ and $p$ are the densities of Mn and holes, respectively, $\langle S^{z}(t) \rangle$ is the average Mn spin, $s_{0}$ is the instantaneous equilibrium value of the average hole spin (determined by $T_{\text{h}}$ and the band splitting $\Delta \! \sim \! \langle S^{z}(t) \rangle$), and $\tau_{\text{sr}}$ is the hole spin relaxation time. The first term on the right hand side is due to the $p$-$d$ spin flip scattering, and it is responsible for the dynamical polarization of holes. If the second term was not efficient enough (if $\tau_{\text{sr}}$ was too long), the holes would get dynamically polarized to such a degree that there would be no more holes with the spin suitable for $p$-$d$ scattering with the Mn spins. Then the demagnetization process would cease.

\begin{figure}
\includegraphics [width=9cm] {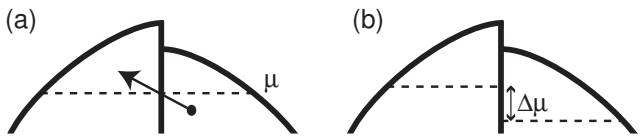}
\caption{Illustration of the spin bottleneck effect for a (III,Mn)V semiconductor. (a) Hole spin-flip scattering event which leads to demagnetization of the localized spins (this is the dominant scattering event when holes are hot and Mn are fully polarized), with $\mu$ being the chemical potential. The band is spin-split by exchange interaction with the localized spins. (b) When the spin relaxation time of holes is finite, such scattering leads to the dynamical polarization of holes, parametrized by the spin splitting $\Delta \mu$ of the quasi-chemical potentials. } \label{fig:bottleneck}
\end{figure}

For antiferromagnetic coupling between the Mn spins and valence band electrons\cite{footnote_beta}
($\beta < 0$ in the standard notation for dilute magnetic semiconductors), spin-flip scattering between hot holes and Mn spins leads to a {\it decrease} of the hole spin polarization, as shown in the schematic illustration in Fig.~\ref{fig:bottleneck}.
Therefore, assuming a finite hole spin relaxation time $\tau_{\text{sr}}$, the net spin in the valence band during the transient dynamics is actually {\em lower} than the equilibrium value corresponding to the instantaneous Mn magnetization.  Let us mention here the existence of a competing process, in which the hot holes are simply relaxing their average spin towards the instantaneous equilibrium value $s_{0}$ [the second term on the left hand side of Eq.~(\ref{eq:dsdt})]. For values of parameters used in our calculations this enhances the effect due to the dynamical polarization.

An important observation is that the assumption of $\theta_{\text{K}}(t) \! \propto \! \langle s^{z}(t) \rangle$ leads to a natural explanation of the temperature dependence of the oveshoot feature in Fig.~\ref{power}(b). At temperatures slightly below $T_{\text{c}}$, the initial Mn polarization is very small.
During the demagnetization, the amount of angular momentum transferred from Mn spins to holes is much smaller than at low $T$, and the spin relaxation of holes can effectively suppress their dynamical polarization ($\Delta \mu \! \approx \! 0$ in Fig.~\ref{fig:bottleneck}). The changes of $\langle s^{z}(t) \rangle$ due to relaxation towards $s_{0}(T_{\text{h}},\Delta)$ are also suppressed at small values of $\Delta$. Thus, the disappearance of the overshoot before the vanishing of the real demagnetization can be qualitatively explained in this scenario. It is not so if we assume that the overshoot simply comes from strong bleaching of all the optical transitions, in which case it should remain visible at all $T<T_{\text{c}}$.

\subsection{Calculation of Mn and hole spin dynamics}  \label{sec:theory}
In order to quantitatively model the transient dynamics of the hole population, we solve the equation for dynamics of the average Mn spin $\langle S^{z}(t) \rangle$:
\begin{equation}
\frac{d}{dt} \langle S^{z}(t) \rangle = \sum_{m=-5/2...5/2}  m \frac{d}{dt} \rho_{m}(t) \,\, ,
\end{equation}
where the time-dependence of the diagonal elements of the localized spin density matrix $\rho_{m}$ is given by the rate equations:
\begin{eqnarray}
 \frac{d}{dt} \rho_{m} &=& -(W_{m-1,m} + W_{m+1,m})\rho_{m} \nonumber \\
 && + W_{m,m+1}\rho_{m+1} + W_{m,m-1}\rho_{m-1} \,\, .
\end{eqnarray}
The transition rates $W_{m,m\pm 1}$ depend on the distribution of holes (their temperature and polarization). The full expressions and details of their calculation are given in Ref.~\onlinecite{Cywinski_PRB07}. Here we only need to know that the spin-flip rates $W$ are proportional to the temperature of holes $T_{\text{h}}$.
The equation for the dynamics of $\langle S^{z}(t) \rangle$ is coupled with Eq.~(\ref{eq:dsdt}) for the dynamics of the average hole spin. Finally, the temperature of holes after the end of the pulse is assumed to decay as $T_{\text{h}}(t)\! \sim \!  \exp (-t/\tau_{\text{E}})$ with energy relaxation time $\tau_{\text{E}}$. The fact that $T_{\text{h}}$ decays here towards zero instead of the final hole temperature (which is about  $50-100$~K) is irrelevant for the dynamics during the first couple of picoseconds --- after $T_{\text{h}}$ drops to about $100$ K the demagnetization becomes very slow and it does not matter for our purposes.

\begin{figure}
\includegraphics [width=9cm] {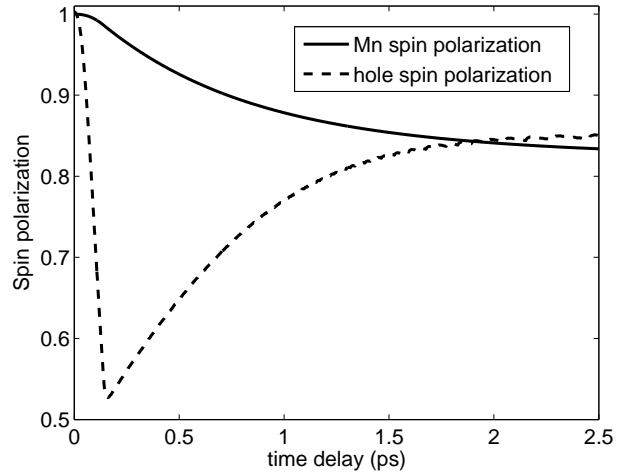}
\caption{Simulation of the time dependence of the Mn spin polarization (the macroscopic magnetization) and the hole spin polarization after excitation with 140~fs-long pulse of light. The influence of the pulse is modeled as a rise of hole temperature  $T_{\rm h}$ from  4~K to 1000~K right after the end of the pulse.  Other parameters are: total $p$ = 10$^{20}$~cm$^{-3}$, energy relaxation time of holes $\tau_{\text{E}}$ = 1~ps, and their spin relaxation time
$\tau_{\text{sr}}$ = 10~fs, $p$-$d$ exchange constant $N_{0}\beta$ = $-1$~eV (with $N_{0}$ being the cation density), which for Mn concentration $x$ = 0.26 leads to the initial spin splitting of the valence band of 70~meV.} \label{fig:theory}
\end{figure}

We have performed calculations using the effective Hamiltonian model,\cite{Dietl_PRB01,Jungwirth_RMP06} in which the mean-field spin splitting is added to the six band Luttinger model. We have used the spin relaxation time $\tau_{\text{sr}}\! = \! $ 10 fs, and adjusted the energy relaxation time $\tau_{\text{E}} \! = \! 1$ ps to fit the observations.
Figure~\ref{fig:theory} shows our calculated time-dependence of Mn and hole spin polarizations after excitation.
The Mn spins demagnetize by $\sim$20\%, in qualitative agreement with the experimental data in Fig.~2, and the time-scale of the dynamical process (demagnetization of Mn and dynamical polarization of holes) is limited by the energy relaxation time $\tau_{\text{E}}$ ---  as $T_{\rm h}$ decreases, the phase space available for hole scattering diminishes, and the efficiency of hole-Mn spin-flip drops.
The depolarization of holes exhibits a prominent overshoot, suggestively similar to the observed MOKE signal.  Thus, under our assumption that $\theta_{\rm K} \! \propto \! \langle s^{z}\rangle$ in GaMnAs, the dynamical polarization of holes can explain the observed temporal profile of the probe response.

\subsection{Energy relaxation of holes}
The distinct overshoot behavior in \nobreak{GaMnAs}, absent in \nobreak{InMnAs}, provides a clue for understanding the ultrafast relaxation dynamics of holes in (III,Mn)V semiconductors.  We attribute the partial recovery of the MOKE signal with a time constant of $\sim$1~ps  to the energy relaxation of highly-excited holes.  It is important to note that much less is known experimentally about the ultrafast dynamics of holes than that of electrons, especially in a
regime of strong excitation.  In our case $\sim10^{20}$~cm$^{-3}$ hot holes are distributed in a very wide energy region (comparable to $k_{\rm B}T_{\rm h}\approx 0.1$~eV, with an effective temperature $T_{\rm h}$ of $\approx 10^3$~K after thermalization).  This regime remains largely unexplored by experiments. Let us reiterate that our $\tau_{\text{E}}$ is the characteristic time-scale of the {\it initial} large energy loss of the very hot holes, i.e., the decay of their temperature from about $10^{3}$~K to $\sim 10^{2}$~K, with the latter value already corresponding to a slow demagnetization dynamics (time-scale of tens of picoseconds).  Theoretical calculations of energy loss rates for high-energy holes suggest that such a highly-excited population will lose a significant part of its energy through phonon emission during the first picosecond,\cite{Woerner_PRB95} in agreement with our measurements.  Moreover, the picosecond energy relaxation of holes explains the rapid termination of the demagnetization process very well.

\section{Conclusions}
We have observed photoinduced sub-picosecond demagnetization in \nobreak{GaMnAs}.  Because of the disordered nature of \nobreak{GaMnAs}, the midinfrared pump strongly excites the valence band.  This leads to demagnetization of the Mn spins, which proceeds as long as the holes are hot.  We propose that the initial overshoot in the drop of the probe Kerr signal is related to a strong excitation of holes near the Fermi level, which bleaches the magneto-optical response of the probe. A possible scenario is the appearance of transient dynamical polarization of hot holes (due to enhanced spin-flip scattering with Mn spins). Under assumption that the probe signal, at frequency of the transitions originating close to the Fermi level, is dominated by average spin polarization in the valence band (and not the average Mn polarization), such a dynamical polarization leads to an overshoot in qualitative agreement with the observation. However, more work on understanding the equilibrium and nonequilibrium magneto-optical properties of \nobreak{GaMnAs} is necessary in order to verify this interesting scenario.
The subsequent sub-picosecond recovery of Kerr rotation is interpreted as a signature of fast energy relaxation of holes.  As expected, ultrafast demagnetization stops after this relaxation time.  Hole spin relaxation time much shorter than the pulse width is inferred from the polarization dependence, consistent with our previously proposed theoretical model.

\begin{acknowledgments}
This work was supported by DARPA (MDA972-00-1-0034), NSF
(DMR-0134058, DMR-0325474, DMR-0325599, and OISE-0530220), and ONR
(N000140410657).
\end{acknowledgments}



\end{document}